\documentclass[3p,twocolumn]{elsarticle} 
\usepackage[latin1]{inputenc}         

\usepackage{amsmath}
\usepackage{units}

\journal{Computer Physics Communications (submitted to proceedings CCP2010)}

\begin{document}

\hyphenation{arc-length}

\newcommand{\vekt}[1]{\mathbf{#1}}
\newcommand{\vekts}[1]{\pmb{#1}}
\newcommand{\uklamm}[2]{\underset{#2}{\underbrace{#1}}}
\newcommand{\kalman}[0]{K\'alm\'an }
\newcommand{\twodim}[0]{two dimensional }
\newcommand{\millepede}[0]{Millepede}
\newcommand{\millepedetwo}[0]{Millepede-II}
\newcommand{\mille}[0]{Mille}
\newcommand{\pede}[0]{Pede}

\begin{frontmatter}

\author[unihh]{Volker Blobel}
\ead{blobel@mail.desy.de}
\author[desyhh]{Claus Kleinwort}
\ead{claus.kleinwort@desy.de}
\author[psieth]{Frank Meier\corref{cor}}
\ead{frank.meier@psi.ch}
\author{on behalf of the CMS collaboration}
\cortext[cor]{Corresponding author}
\address[unihh]{Institut f\"ur Experimentalphysik, Universit\"at Hamburg, Germany}
\address[desyhh]{DESY Deutsches Elektronen-Synchrotron, Notkestra\ss e 85, 22607 Hamburg, Germany}
\address[psieth]{Paul~Scherrer~Institut, 5232~Villigen, Switzerland and ETH~Zurich, 8093~Z\"urich, Switzerland}
\title{Fast alignment of a complex tracking detector using advanced track models}
\date{}
\begin{abstract}
The inner silicon detector of the Compact Muon Solenoid experiment (CMS) at CERN's LHC consists of 16\,588 modules. Charged-particle tracks in the detector are used to improve the accuracy to which the position and orientation of the modules is known. This contribution focuses on the \millepedetwo{} algorithm, one of the two alignment algorithms used by CMS.  Recently an advanced track model has been introduced into the CMS alignment procedure, which is based on the ``Broken Lines'' model and is able to take multiple Coulomb scattering in the detector material properly into account. We show the unique approach needed for solving the alignment problem in a reasonable amount of time. Emphasis is given to the mathematical treatment of the problem.

\end{abstract}
\begin{keyword}
Particle tracking detectors (Solid-state detectors) \sep Alignment  \sep \millepede{} \sep Broken Lines \PACS 29.40.Wk \sep Solid-state detectors \PACS 06.20.fb \sep Standards and calibration 
\end{keyword}
\end{frontmatter}





\graphicspath{{./img/}}



\newcommand{\degree}[0]{$^\circ$}


\section{Introduction}
The inner tracker of the CMS \cite{cmsjinst} detector at CERN's \emph{Large Hadron Collider} (LHC) is designed to track the paths of charged particles as they pass through the detector. In order to determine the parameters of such tracks accurately, the positions and orientations of the detector modules need to be known to at least their intrinsic resolution ($\approx10\,\mu m$).

%

\paragraph{Track-based alignment}
Track-based alignment is a least squares minimization problem ($\chi^2$ minimization). The one or more dimensional residuals, $\vekt{r}_{ij}$, for a given hit $i$ on a given track $j$ represent the distance between the hit 
location predicted by the track model (helix or other suitable model) and the actual hit measurements from the detector modules, calculated using the knowledge of the geometry before the alignment. Using the covariance 
matrix $\vekt{V}_{ij}$ of each measurement, the global $\chi^2$ is calculated as:

\small \begin{equation} \label{eqn.chi2ali} \chi^2(\vekt{p},\vekt{q}) 
	= \sum_j^\text{tracks} \sum_i^\text{hits} \vekt{r}_{ij}^T(\vekt{p},\vekt{q}_j) \,\vekt{V}_{ij}^{-1}\, \vekt{r}_{ij}(\vekt{p},\vekt{q}_j)
\end{equation} \normalsize
and minimized with respect to the alignment parameters $\vekt{p}$ and the track parameters $\vekt{q}$ of all tracks. 
Millions of tracks are required for accurate parameter determination. 
As the effects of multiple scattering in the detector material can be significant in comparison with the intrinsic measurement resolution, they have to be described properly.

\section{\millepedetwo}
Track-based alignment with the \millepedetwo{} algorithm \cite{blobel06mp} uses a simultaneous fit of a large number of tracks to determine corrections $\Delta\vekt{p}$ to the positions and orientations of the detector components. It assumes 
independent and uncorrelated scalar measurements $m_{ij}$ with errors $\sigma_{ij}$. 
Using block-matrix theorems the alignment problem can be reduced (without approximations)
to a matrix equation $\vekt{A}_\text{align}\Delta\vekt{p}=\vekt{b}_\text{align}$ of the size of the number of alignment parameters $n_\text{align}$. 
This reduction requires the covariance matrices of all individual track fits to construct $\vekt{A}_\text{align}$. 
The track model $m_{ij}=f_{ij}(\vekt{p},\vekt{q}_j)$  is linearized at some starting values of the alignment and track parameters:
\begin{multline}
 r_{ij} = m_{ij} -f_{ij}(\vekt{p}_0,\vekt{q}_{j0})  -\frac{\partial f_{ij}}{\partial\vekt{p}}\Delta\vekt{p}-\frac{\partial f_{ij}}{\partial\vekt{q}_{j}}\Delta\vekt{q}_{j}\\
\chi^2(\Delta\vekt{p},\Delta\vekt{q})=
  \sum_j^\text{tracks}\sum_i^\text{meas.}
  \frac{ r_{ij}^2}{\sigma_{ij}^2}
\end{multline}
Linearization  and outlier downweighting or rejection require an iterative procedure. To setup the matrix equation of alignment parameters and update it during iterations, 
for each track a fit is performed to determine with the current values of the alignment parameters $\vekt{p}=\vekt{p}_0+\Delta\vekt{p}$ the residuals $r_{ij}$ and at least in the first iteration the covariance matrix $\vekt{V}(\vekt{q}_j)$ of all track parameters along the trajectory.
Constraints from the physical structure of the problem are treated via Lagrange multipliers.  The matrix  $\vekt{A}_\text{align}$ is usually sparse (not all detector components are connected by tracks). The equation system can 
be solved with different methods. With simple \emph{matrix inversion} the solution is $\Delta\vekt{p}=\vekt{A}_\text{align}^{-1}\vekt{b}_\text{align}$. The calculation of $\vekt{A}_\text{align}^{-1}$ is done once in the first iteration. 
As inversion requires ${\cal O}(n_\text{align}^3)$ operations and the storage of the full matrix this method is slow and memory consuming,  but 
provides the covariance matrix of the solution. 
With typical computing resources the practical limit for matrix inversion is at $n_\text{align}$ of a few $10^4$.
Alternatively the fast iterative MINRES algorithm \cite{minres} can be used. It minimizes $\Vert\vekt{A}_\text{align}\Delta\vekt{p}-\vekt{b}_\text{align}\Vert^2$ with $n_\text{it}$ 
internal iterations using ${\cal O}(n_\text{align}^2 n_\text{it})$ operations. It needs only the sparse representation (the non zero elements) of the matrix saving a large fraction of memory space.

In the technical implementation \millepedetwo{} is split into two parts: The first (\emph{\mille}) is integrated in the track reconstruction software and writes for each track a record to a special binary file. A record contains for all 
measurements of that track the residuum, its error and the non-zero derivatives versus the alignment and track parameters. The second (\emph{\pede}) is a standalone program implemented in FORTRAN reading the binary files and 
performing the solution according to the specified steering.
  
\section{Broken lines}
Charged particles traversing layers of material experience \emph{multiple scattering}, mainly due to Coulomb interaction with the electrons in the atoms, resulting in random changes in direction and spatial position  
 with expectation values of zero and variance depending on the traversed material and the particle momentum. The broken lines model is a track refit adding the description of multiple scattering to an initial trajectory and able 
 to determine the covariance matrix of all track parameters. Therefore it is well suited as track fit for \millepedetwo{}. 
The original formulation \cite{blobel06bl} describes the case of a tracking detector with a solenoidal magnetic field and independent \twodim tracking in the bending and perpendicular to the bending plane. It constructs the 
planar trajectories from the measurements including the material around those as thick scatterers.  

\begin{figure}
\begin{center}
\includegraphics[width=1.0\linewidth]{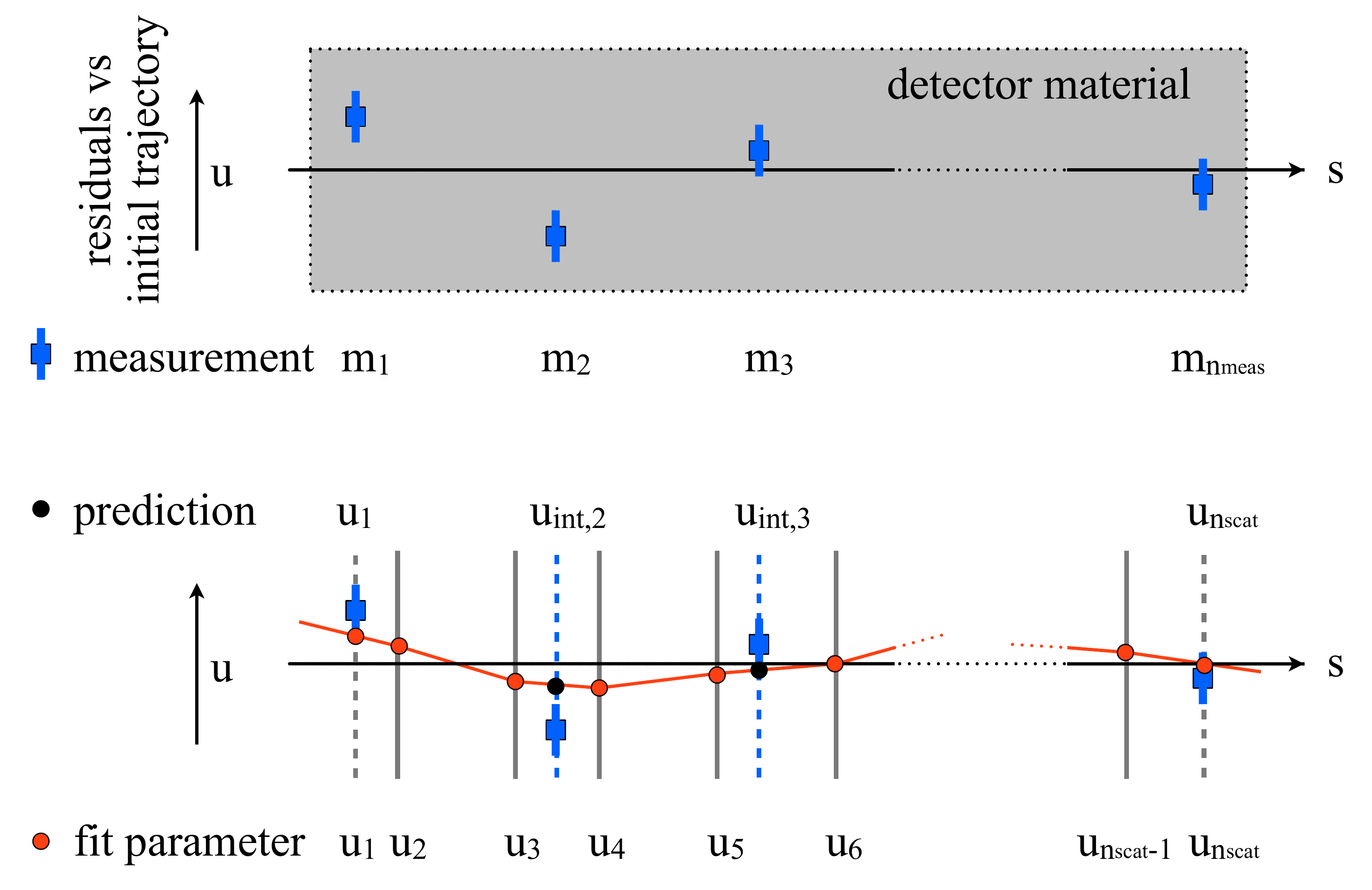}
\end{center}
\caption{Simple example in one plane, no curvature correction $\Delta\kappa$, measurements $m_i$ in planes perpendicular to the track direction and homogeneous material distribution.
\emph{Top:} Residuals versus initial trajectory along arc-length $s$.
\emph{Bottom:} Broken lines trajectory based on thin scatterers with offsets $u$.  The material between measurement $i$ and $i+1$ is described with two thin scatterers at $s=(s_{i+1}+s_{i})\pm(s_{i+1}-s_{i})
/\sqrt{12}$. First and last measurement define additional offsets.
The fit  prediction $u_{\text{int},i}$ for the measurement $m_i$ is obtained by interpolation between the enclosing scatterers: $u_{\text{int},i}=f(u_{2i-1},u_{2i}),\:(u_{\text{int},1}=u_1,\:u_{\text{int},n_{\text{meas}}}=u_{n_{\text{scat}}})$.}
\label{fig.brl}
\end{figure}

In the presence of  measurements with components in both planes  a single trajectory in space is required. 
In the following this is described for a track with $n_\text{meas}$ one- or two- dimensional measurements in $n_\text{plane}$ planes.
The material between adjacent measurements planes is represented
by up to two thin scatterers (zero thickness) with the same mean and variance of the material distribution. A thin scatterer produces no direct spatial shifts but only a \twodim scattering angle $\vekts{\beta}$ with a variance $\vekt{V}_
{\beta}=\theta_0^2\, \bigl( \begin{smallmatrix} 1 & 0 \\  0 & 1  \end{smallmatrix} \bigl)$  \cite{pdg}.  The trajectory is constructed now from the thin scatterers adding the measurements by interpolation of the enclosing scatterers. At each thin scatterer a \twodim offset  $\vekt{u}=(x_\perp,y_\perp)$ in
 the \emph{curvilinear frame} (perpendicular to the track direction) is defined as fit parameter of the track together with a common curvature correction $\Delta\kappa$.
To define the start and end of the trajectory  the first and last measurement planes are added to the sequence of scatterers.
The number of scatterers $n_\text{scat}$ is between two (no material at all) and $2\,n_\text{plane}$ (thick scatterers between all planes) and the number of track parameters to be fitted is $n_\text{par}=2\,n_\text{scat}+1$. Figure \ref{fig.brl} illustrates this construction.
 
With small distortions $\Delta\vekt{u}$ of the offset, $\Delta\vekts{\alpha}$ of the slope and $\Delta\kappa$ of the curvature as local track parameters the offsets propagate like:
\begin{multline}
 \Delta\vekt{u}_{i+1} = \frac{\partial\vekt{u}_{i+1}}{\partial\vekt{u}_i}\Delta\vekt{u}_i+ \frac{\partial\vekt{u}_{i+1}}{\partial\vekts{\alpha}_i}\Delta\vekts{\alpha}_i+ \frac{\partial\vekt{u}_{i+1}}{\partial\kappa}\Delta\kappa\\
 =\vekt{J}_i\Delta\vekt{u}_i+\vekt{S}_i\Delta\vekts{\alpha}_i+\vekt{d}_i\Delta\kappa
\end{multline}

The derivatives $\vekt{J}$,  $\vekt{S}$ and $\vekt{d}$ can be obtained from the curvilinear Jacobian and the transformation between the local and curvilinear track parameters:
\begin{multline}
 \frac{\partial(\kappa,\vekts{\alpha},\vekt{u})_{i+1}}{\partial(\kappa,\vekts{\alpha},\vekt{u})_{i}}=
 \frac{\partial(\kappa,\vekts{\alpha},\vekt{u})_{i+1}}{\partial(q/p,\lambda,\phi,x_\perp,y_\perp)_{i+1}}\\
 \frac{\partial(q/p,\lambda,\phi,x_\perp,y_\perp)_{i+1}}{\partial(q/p,\lambda,\phi,x_\perp,y_\perp)_{i}}
 \frac{\partial(q/p,\lambda,\phi,x_\perp,y_\perp)_{i}}{\partial(\kappa,\vekts{\alpha},\vekt{u})_{i}}
\end{multline}
 The case of a constant magnetic field is described in \cite{strandlie06}. Solving for the slope correction yields:
\begin{equation}
 \Delta\vekts{\alpha}_i=\vekt{S}_i^{-1}( \Delta\vekt{u}_{i+1}-\vekt{J}_i\Delta\vekt{u}_i-\vekt{d}_i\Delta\kappa)
\end{equation}
With a triplet of three adjacent offsets ($\vekt{u}_{-},\vekt{u}_{0},\vekt{u}_{+}$) two slopes can be determined at $\vekt{u}_0$:
\begin{multline}
 \vekts{\alpha}_+=\vekt{W}_+( \vekt{u}_+-\vekt{J}_+\vekt{u}_0-\vekt{d}_+\Delta\kappa), \vekt{W}_+=\vekt{S}_+^{-1} \\
 \vekts{\alpha}_-=\vekt{W}_-( \vekt{J}_-\vekt{u}_0-\vekt{u}_-+\vekt{d}_-\Delta\kappa), \vekt{W}_-=-\vekt{S}_-^{-1}
\end{multline}
The difference $\vekts{\beta}= \vekts{\alpha}_+- \vekts{\alpha}_-$ measures the change of direction due to multiple scattering at the central scatterer ($\vekt{u}_0$):
\begin{multline}
 \vekts{\beta}=\vekt{W}_+\vekt{u}_+-(\vekt{W}_+\vekt{J}_++\vekt{W}_-\vekt{J}_-)\vekt{u}_0+\vekt{W}_-\vekt{u}_-\\
 -(\vekt{W}_+\vekt{d}_++\vekt{W}_-\vekt{d}_-)\Delta\kappa
\end{multline}
In case of a measurement instead of a scatterer at $\vekt{u}_0$ there is no scattering angle and solving the previous equation for $\vekt{u}_\text{int}=\vekt{u}_0$ with $\vekts{\beta}\equiv\vekt{0}$ using $\vekt{N}=(\vekt{W}_+\vekt{J}_++\vekt{W}_-\vekt{J}_-)^{-1}$ results in the interpolation equation:
\begin{multline}
 \vekt{u}_\text{int}=\vekt{N}(\vekt{W}_+\vekt{u}_++\vekt{W}_-\vekt{u}_-)\\ -\vekt{N}(\vekt{W}_+\vekt{d}_++\vekt{W}_-\vekt{d}_-)\Delta\kappa
\end{multline}
The interpolated offsets $\vekt{u}_{\text{int},i}$ and the scattering angles $\vekts{\beta}_i$ are predictions depending linear on the fit parameters $\vekt{q}=(\Delta\kappa$,$\vekt{u}_1,..,\vekt{u}_{n_{scat}}$). Their expectation values 
are the \emph{measured} values from the initial trajectory: $\vekt{P}\cdot\langle\vekt{u}_{int,i}\rangle=\vekt{m}_i $ (residual to initial trajectory) and $\langle\beta_i\rangle=0$.  $\vekt{P}=\frac{\partial\vekt{m}}{\partial\vekt{u}}$ is 
the projection matrix into the measurement system, where the covariance matrices $V_{\text{meas},i}$, are diagonal. The fit parameters are determined by minimizing:
\begin{multline}
\chi^2(\vekt{q})
   = \sum_{i=2}^{n_\text{scat}-1} \vekts{\beta}_i(\vekt{q})^T \vekt{V}^{-1}_{\beta,i} \vekts{\beta}_i(\vekt{q})\:+\\
    \sum_{i=1}^{n_\text{meas}}(\vekt{m}_i-\vekt{P}_i\vekt{u}_{\text{int},i}(\vekt{q}))^T \vekt{V}^{-1}_{\text{meas},i} (\vekt{m}_i-\vekt{P}_i\vekt{u}_{\text{int},i}(\vekt{q}))
\end{multline}  
 
The minimization leads to a linear equation system $\vekt{A}\vekt{q}=\vekt{b}$ of size $n=n_\text{par}$. As all $\vekt{u}_{\text{int},i}$ depend only on two and all $\vekts{\beta}_i$ on three adjacent offsets (and all 
 on $\Delta\kappa$), the matrix $\vekt{A}$ is a bordered band matrix with border size $b=1$  (curvature part) and band width $m=3\cdot\text{dimension}(\vekt{u})-1=5$ (offset part): $\vekt{A}_{kl}=0$ for $\text{min}(k,l)>b$ and $\text{abs}(k-l)>m$.
 This special structure of the matrix allows the usage of the fast
 root-free Cholesky decomposition $\vekt{A}_u=\vekt{L}\vekt{D}\vekt{L}^T$ of the band part of the matrix into a diagonal matrix $\vekt{D}$ and a left unit triangular matrix $\vekt{L}$ to obtain the solution with ${\cal O}(n(m+b)
^2)$ and the full covariance matrix with ${\cal O}(n^2(m+b))$ operations (instead of simple inversion using $O(n^3)$). 

\millepedetwo{} has been updated to check for each track fit the matrix $\vekt{A}$ for a bordered band structure to benefit in this case from the fast solution by root-free Cholesky decomposition.

\section{CMS silicon tracker alignment}
Track-based alignment with \millepedetwo{} and broken lines is one of the methods used for alignment of the CMS silicon tracker.

For detailed alignment of the CMS silicon tracker consisting out of 16\,588 modules several $10^4$ parameters are determined with \millepedetwo{} using MINRES from millions of tracks. The \emph{\mille} step for the creation of binary files is performed 
by many jobs in parallel, each processing different parts of the input data (tracks) in typically one hour. All binary files are processed by a single \emph{\pede} step to fit the alignment parameters.  The time spent for the solution can be
significant (many hours). The MINRES algorithm needs typically several hundred internal iterations and is therefore about a $100$ faster than simple matrix inversion. The most time consuming part in MINRES is the product of a 
symmetric matrix and a vector of size $n_\text{align}$. This has been parallelized with OpenMP$^\text{TM} $\cite{openmp} to run \emph{\pede} on multiple cpu cores. 

In the CMS tracking model measurement and multiple scattering happen at the same point: the silicon sensor layers (as thin scatterers). Therefore the number of scatterers is equal to the number of silicon layers 
($\,n_\text{scat}=\,n_\text{plane}$) and interpolation is trivial ($\vekt{u}_{\text{int},i}=\vekt{u}_i$).

For a constant axial magnetic field $\vekt{B}=(0,0,B_z)$ in the limit of small curvature the derivatives for the track propagation from scatterer $i$ to $i+1$ are simple functions of the arc-length $s$ along the track:
\begin{multline}
 \vekt{J}=\begin{pmatrix} 1 & 0 \\ 0 & 1 \end{pmatrix},\; \vekt{S}=\Delta s_i\begin{pmatrix} 1 & 0 \\  0 & 1  \end{pmatrix}, \\ \;\vekt{d}=\tfrac{1}{2}(\Delta s_i \sin\vartheta)^2 \begin{pmatrix} 1 \\ 0 \end{pmatrix}
\end{multline}
Here $\vartheta$ is the angle between track and magnetic field direction and  $\Delta s_i =s_{i+1}-s_i=1/\delta_i$. The scattering angles $\vekts{\beta}_i$ are in this case (similar to equations (4) and (7) in \cite{blobel06bl}) simply:

 \begin{multline}
\vekts{\beta}_i = \left(\vekt{u}_{i-1}\delta_{i-1}
	         -\vekt{u}_i(\delta_{i-1}+\delta_i)
		 +\vekt{u}_{i+1}\delta_{i}\right) \\
              - \tfrac{1}{2}\sin\vartheta^2(\Delta s_{i-1} + \Delta s_i)  \begin{pmatrix} 1 \\ 0 \end{pmatrix}\Delta\kappa
\end{multline}

This is the first implementation of the broken lines model in three dimensions. The approximated diagonal forms of $\vekt{J}$ and $\vekt{S}$ lead to a reduced  band width  $m=4$ providing an even faster track fit and due to fewer non-zero derivatives to smaller \millepede{} binary files. This implementation is now regularly used for CMS track-based alignment with \millepedetwo{}. 


\section{Results}
\begin{figure}
\begin{center}
\includegraphics[width=0.90\linewidth]{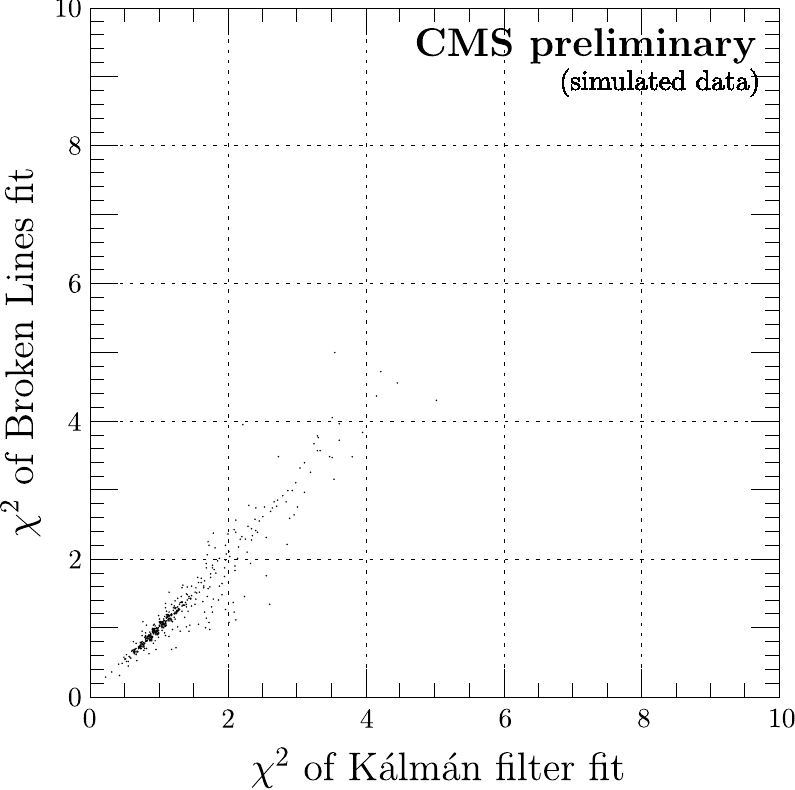}

\bigskip

\includegraphics[width=0.92\linewidth]{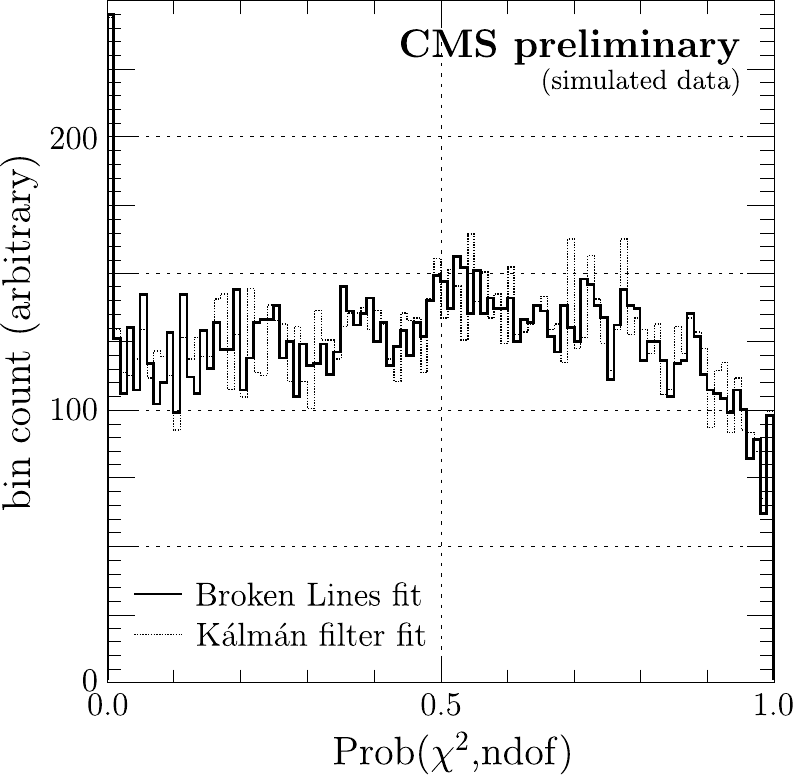}
\end{center}
\caption{\emph{Top:} Correlation of $\chi^2$ values of single tracks, comparing the broken lines fit with the \kalman filter fit. 
\emph{Bottom:} Probability values of the $\chi^2$ values of single tracks.}
\label{fig.chi2}
\end{figure}

\paragraph{Equivalence of track models} The reference algorithm for tracking is the \kalman filter algorithm\cite{fruehw87}, a sequential track fitting algorithm stepping between scatterers by adding measurements and process noise, without calculating the full covariance matrix. Equivalence is shown in figure \ref{fig.chi2}, both from track fits (12\,000 simulated ``isolated muon'' tracks). The correlation is good and no differences between the two fits are visible in the probability distributions, even as a function of track parameters (momentum, track angle, not shown here).

\paragraph{Speed performance}
The performance of the Cholesky decomposition was measured to be 7 times faster than full inversion of the matrix in the track fit (subset of 250\,000 cosmic tracks).

Typical alignment of the full detector using $4.5\cdot10^ 6$ tracks and solving for 57\,000 parameters takes 6 hours (one node, up to 8\,GB memory, solution step (\emph{\pede}) only). Parallelization of some parts of the code (OpenMP$^\text{TM}$) improves speed on average by a factor of 3 (7 cores used). 

\section{Conclusions}
The use of a suitable track model for alignment and its equivalence to the standard \kalman filter fit has been shown.
A simple extension to \millepedetwo{} (Cholesky decomposition for the internal track fit) allowed for performance optimization. Parallelization improves speed by a factor of 3.



\end{document}